\documentclass[a4paper,11pt]{article}
\usepackage{jheppub} 
\usepackage{csquotes}
\usepackage{hyperref}

\arxivnumber{2604.05262} 

\title{\boldmath Residual Symmetries and Their Algebras in the Kerr-Schild Double Copy}

\author{B. P. Holton}
\affiliation{Durham University,\\
Department of Mathematical Sciences,\\
Upper Mountjoy Campus, Stockton Road, Durham, DH1 3LE, UK}

\emailAdd{brandonholton557@gmail.com}

\abstract{The Kerr-Schild double copy (KSDC) is well-known for relating exact classical solutions between Yang-Mills theory and theories of gravity. However, whether this correspondence provides a more fundamental mapping between the underlying symmetries of gauge theory and gravity remains an underdeveloped area of research in the contemporary double copy program.
\\ \\
In this paper, we demonstrate that the KSDC correspondence does not provide a mapping between the residual symmetry structures of the Kerr-Schild ansatz in Yang-Mills theory and gravity. On the gauge theory side, residual symmetries form an infinite-dimensional algebra of functions along null directions. On the gravitational side, residual diffeomorphisms preserving the Kerr-Schild form of the Schwarzschild metric generate a conformal algebra on $S^2$, which decomposes into Killing vectors and proper conformal Killing vectors (CKVs). While the Killing sector reproduces the expected global isometries, the CKV sector yields an infinite-dimensional algebra after imposing asymptotic flatness and horizon regularity.
\\ \\
This appears to contradict the fact that the Schwarzschild solution admits no proper conformal symmetries. We resolve this apparent contradiction by constructing a Weyl-compensated BRST complex, showing that the CKV sector is BRST-exact and therefore trivial in cohomology, so that the physical symmetry algebra reduces to the global isometries of Schwarzschild. This demonstrates that the KSDC introduces an enlarged symmetry structure at the level of the ansatz, but preserves physical symmetries after a cohomological reduction, revealing a fundamental mismatch between Yang-Mills and gravity at the level of residual symmetries.}

\begin{document}
\maketitle
\flushbottom

\section{Introduction}

The relationship between gauge theory and gravity revealed by the double copy has led to remarkable insights into the structure of classical and quantum field theories. Originally formulated in the context of scattering amplitudes \cite{k, l, m, n, t, u}, the double copy has since been extended to a variety of classical settings, including exact solutions constructed using the Kerr-Schild (KS) ansatz \cite{a, b, y, cc, dd, ff}. In this framework, solutions of Yang-Mills theory can be mapped to solutions of general relativity through a linearization of the Einstein equations, providing a concrete realization of the correspondence at the level of fields.
\\ \\
While much of the existing literature has focused on the mapping between solutions, comparatively less attention has been paid to the relationship between the \textit{symmetry structures} underlying these solutions. In particular, it is not clear to what extent the double copy preserves residual symmetries -- transformations that leave a given ansatz invariant. This question is especially natural in the Kerr-Schild setting, where the ansatz imposes a rigid geometric structure that can admit nontrivial symmetry enhancements \cite{g, w}. Recent work has begun to explore symmetry aspects of the double copy, in particular, the convolutional \cite{p, q, x, aa, bb} and self-dual \cite{o} sectors. Symmetries and applications in $\mathcal{N}=2$ supergravities have been studied as well \cite{c, d, e, f}. Related work has also highlighted the appearance of diffeomorphism and kinematic algebra structures in classical double copy constructions and associated gauge-theoretic systems \cite{fstar}. Nevertheless, a systematic understanding of residual symmetry structures in the Kerr-Schild double copy remains incomplete.
\\ \\
In this paper, we investigate residual symmetry structures of Yang-Mills theory and gravity in the Kerr-Schild double copy for the Schwarzschild solution. On the gauge theory side, residual transformations preserving the Kerr-Schild form of the gauge field are found to be arbitrary functions along null directions, forming an infinite-dimensional current algebra. On the gravitational side, residual diffeomorphisms preserving the Kerr-Schild form of the Schwarzschild metric give rise to a conformal symmetry algebra on the two-sphere. This algebra decomposes into two distinct sectors: a finite-dimensional Killing sector corresponding to global isometries, and an infinite-dimensional sector generated by proper conformal Killing vectors (CKVs). In contrast to approaches centered on kinematic algebras associated with interaction structures or color-kinematics duality (see ref. \cite{fstar}), our analysis focuses on residual symmetry algebras preserving the Kerr-Schild ansatz itself and their subsequent BRST cohomological reduction.
\\ \\
At first sight, the appearance of an infinite-dimensional CKV sector appears to contradict the well-known fact that Schwarzschild spacetime admits no proper conformal symmetries. The resolution of this apparent discrepancy is one of the main results of this work. We show that, although these conformal symmetries arise naturally at the level of the Kerr-Schild ansatz, they do not correspond to physical symmetries of the spacetime. Instead, they are removed upon passing to BRST cohomology \cite{i, j, z}.
\\ \\
To demonstrate this, we construct a unified, Weyl-rescaled BRST complex associated with the residual symmetry algebra. The Killing sector is shown to reproduce the standard BRST complex for global symmetries, corresponding to the Chevalley-Eilenberg cohomology of the isometry algebra. In contrast, the proper CKV sector does not admit a standard realization on the space of fields, as it acts on the metric only up to Weyl rescalings. We show that this obstruction can be resolved by introducing a minimal Weyl compensator, allowing the CKV transformations to be realized on an extended field space. The resulting sector forms a BRST-contractible subcomplex, and is therefore cohomologically trivial. Consequently, the full BRST cohomology reduces to the finite-dimensional isometry algebra of Schwarzschild spacetime.
\\ \\
This provides a cohomological interpretation of the symmetry structure of the Kerr-Schild double copy: while the ansatz introduces an enlarged residual symmetry algebra, the additional degrees of freedom correspond to redundancies rather than physical symmetries. In particular, the double copy does not induce a direct correspondence between the full residual symmetry algebras of gauge theory and gravity, but instead preserves the physical symmetry content only after an appropriate cohomological reduction.
\\ \\
This paper both unifies and formalizes the results of \cite{a*, b*}, clarifying several points and presenting them in a more streamlined framework. The paper is organized as follows: In Section 2, we derive the residual gauge transformations of Yang-Mills theory preserving the Kerr-Schild ansatz and determine the associated current algebra. In Section 3, we classify the residual diffeomorphisms preserving the Kerr-Schild form of the Schwarzschild metric, and show that they decompose into Killing and proper CKV sectors. In Section 4, we construct the BRST complex for these symmetries, demonstrate the cohomological triviality of the CKV sector, and show that the physical symmetry algebra is recovered as the BRST cohomology. We conclude in Section 5 with a discussion of the implications for the double copy and possible extensions of this work.

\section{Yang-Mills Residual Symmetries}

We begin by classifying the residual gauge transformations of Yang-Mills theory that preserve the Kerr-Schild (KS) ansatz in Schwarzschild spacetime. Throughout this paper, we work with mostly-plus signature $(-,+,+,+)$ and take the background metric to be flat Minkowski space $\eta_{\mu \nu}$.
\\ \\
We also adopt spherical Minkowski coordinates $(t,r,\vartheta,\phi)$, so that the null vector $k_\mu$ can be written as

\[
k_\mu = (-1, 1, 0, 0) ~~~~~,~~~~~ k^\mu = (1, 1, 0, 0),
\]
\\
where $k^\mu k_\mu = 0$, as required by the null condition.

\subsection{Derivation of Residual Symmetries for a Non-Abelian Field}

Consider a non-Abelian Yang-Mills field $A_\mu^a(x)$ with Lie algebra $\mathfrak{g}$ and structure constants $f^{abc}$, where Latin indices $a,b,c$ label generators in the adjoint representation. The KS ansatz for the field is

\begin{equation}
   \label{1} A_{\mu}^a(x) := \Phi^a(x) k_\mu,
\end{equation}
\\
where $\Phi^a(x)$ is a scalar profile. Under an infinitesimal gauge transformation with parameter $\Lambda^a(x)$, the gauge field transforms as

\begin{equation}
   \label{2} \delta_\Lambda A_{\mu}^a(x) = \partial_\mu \Lambda^a(x) + g f^{abc} A_\mu^b(x) \Lambda^c(x)
\end{equation}
\\
with Yang-Mills coupling $g$. We require that the transformed field $A_{\mu}^{a'}(x)$ preserves the KS form, \eqref{1}, so that

\begin{equation}
  \label{3} A_{\mu}^{a'}(x) \stackrel{!}{=} A_\mu^a(x) + \delta_\Lambda A_\mu^a(x) = [\Phi^a(x) + \delta_\Lambda \Phi^a(x)] k_\mu.
\end{equation}
\\
Substituting \eqref{1} into \eqref{2} and comparing with \eqref{3} yields

\begin{equation}
   \label{4} \delta_\Lambda \Phi^a(x) k_\mu = \partial_\mu \Lambda^a(x) + g f^{abc} \Phi^{b}(x) k_\mu \Lambda^c(x).
\end{equation}
\\
Since the left-hand side of \eqref{4} is proportional to $k_\mu$, the right-hand side must also be proportional to $k_\mu$. Contracting both sides with $k^\mu$ and using $k^\mu k_\mu = 0$ yields the constraint

\begin{equation}
  \label{5} k^\mu \partial_\mu \Lambda^a(x) = 0.
\end{equation}
\\
This first-order PDE is solved by the method of characteristics. In spherical Minkowski coordinates, $k^\mu \partial_\mu = \partial_t + \partial_r$, since the angular components of $k^\mu$ vanish. Thus, \eqref{5} implies that the gauge parameters are constant along outgoing null rays parameterized by retarded time $u := t-r$. The general solution takes the form

\begin{equation}
   \label{6} \Lambda^a(t,r) = f^a(u),
\end{equation}
\\
where $f^a(u)$ are arbitrary smooth functions. Note that, in general, there is nothing in PDE \eqref{5} that necessarily prevents \eqref{6} from have an explicit dependence on $(\theta,\phi)$. However, we can exclude this possibility by observing the following: consider the angular components of \eqref{4}:

\begin{equation}
    \delta_\Lambda \Phi^a k_A = \partial_A \Lambda^a(x) + g f^{abc} \Phi^b(x) \Lambda^c(x) k_A.
\end{equation}
\\
Now, the left-hand side vanishes since $k_A = 0$ in spherical Minkowski coordinates. The color interaction term is proportional to $k_A$, and hence vanishes for the same reason. We are left with

\begin{equation}
    \partial_A \Lambda^a(x) = 0.
\end{equation}
\\
Thus, $\Lambda^a(x)$ must be independent of the angular coordinates by the very construction of the Kerr-Schild ansatz.
\\ \\
Substituting \eqref{6} into \eqref{4} yields the induced transformation on the scalar profile:

\begin{equation}
   \label{7} \delta_f \Phi^a(u) =-f_{,u}^a (u) + g f^{abc} \Phi^b(u) f^c(u),
\end{equation}
\\
where $f_{,u}^a(u)$ denotes the derivative of $f^a(u)$ with respect to $u$.
\subsection{Algebra Generated by Residual Symmetries}

Having established the explicit form of the transformations acting on $\Phi^a(u)$, we now examine the algebra they generate. The residual transformations define linear operators $\delta_f$ acting on the scalar profile $\Phi^a(u)$. These transformations close under commutation:

\begin{equation}
   \label{8} [\delta_f, \delta_h] \Phi^a(u) = \delta_{[f,h]} \Phi^a(u),
\end{equation}
\\
where

\begin{equation}
   \label{9} [f,h]^a(u) := g f^{abc} f^b(u) h^c(u).
\end{equation}
\\
Let $\mathfrak{g}_{\text{res}}$ denote the Lie algebra of transformations $\delta_f$ equipped with bracket \eqref{9}. Identifying each transformation with its parameter function $f^a(u)$, define the linear map

\begin{equation}
   \label{10} \Psi : \mathfrak{g}_{\text{res}} \rightarrow \mathfrak{g} \otimes C^\infty(\mathbb{R}) ~~~~~,~~~~~\delta_f \mapsto f^a(u) T^a,
\end{equation}
\\
where $T^a$ are generators of $\mathfrak{g}$. This map establishes a Lie algebra isomorphism

\begin{equation}
   \label{11} \mathfrak{g}_{\text{res}} \cong \mathfrak{g} \otimes C^{\infty}(\mathbb{R})
\end{equation}
\\
with Lie bracket inherited pointwise from $\mathfrak{g}$. Hence, the scalar  furnishes a representation of $\mathfrak{g}_{\text{res}}$ and the residual symmetry algebra \eqref{11} takes the form of a classical current algebra along outgoing null directions.
\section{Gravitational Residual Symmetries}

We now classify the residual diffeomorphisms that preserve the KS form of the Schwarzschild metric,

\begin{equation}
   \label{12} g_{\mu \nu} := \eta_{\mu \nu} + \varphi(x) k_\mu k_\nu,
\end{equation}
\\
where $\eta_{\mu \nu}$ is the flat Minkowski background, $\varphi(x)$ is the scalar profile, and $k_\mu$ is once again the null vector, which satisfies $k^\mu k_\mu = 0$. In spherical coordinates $(t,r,\vartheta,\phi)$, one may take

\begin{equation}
    \label{13} \varphi(x) := \frac{2GM}{r}
\end{equation}
\\
so that metric \eqref{12} reproduces the exact spacetime geometry of Schwarzschild \cite{r, cc, dd, kk}.

\subsection{Derivation of Kerr-Schild Preservation Condition}

Residual diffeomorphisms are infinitesimal coordinate transformations generated by vector fields $\xi^\mu$ that preserve the Kerr-Schild structure of \eqref{12}. Under such a transformation, the metric varies according to the Lie derivative,

\begin{equation}
   \label{14} \delta_\xi g_{\mu \nu} = (\mathcal{L}_\xi g)_{\mu \nu} := \xi^\rho \partial_\rho g_{\mu \nu} + 2 \partial_{(\mu} \xi^\rho g_{\nu)\rho}.
\end{equation}
\\
Preservation of the KS form requires that this variation be proportional to $k_\mu k_\nu$,

\begin{equation}
   \label{15} \xi^\rho \partial_\rho g_{\mu \nu} + 2 \partial_{(\mu} \xi^\rho g_{\nu)\rho} \stackrel{!}{=} \alpha(x) k_\mu k_\nu,
\end{equation}
\\
where $\alpha(x)$ is a smooth function. Substituting \eqref{12} into Lie derivative \eqref{15} yields,

\begin{equation}
   \label{16} \xi^\rho \partial_\rho \eta_{\mu \nu} + 2 \partial_{(\mu} \xi^\rho \eta_{\nu)\rho} + (\xi^\rho \partial_\rho \varphi) k_\mu k_\nu + 2 \varphi \partial_{(\mu} \xi^\rho k_{\nu)}k_\rho \stackrel{!}{=} \alpha(x) k_\mu k_\nu.
\end{equation}
\\
\textbf{Note:} Although the background is flat, $\partial_\rho \eta_{\mu \nu} \neq 0$ due to the coordinate dependence of the Minkowski metric. This is the trade-off we must consider when we adopt spherical coordinates: the null vector is constant, but the Minkowski background is not. Moreover, we remark that terms proportional to $k_\mu k_\nu$ already preserve the KS form, so we may absorb the scalar variation into a redefinition of $\alpha(x)$. Define

\begin{equation}
   \label{17} \zeta(x) := \alpha(x)-\xi^\rho \partial_\rho \varphi.
\end{equation}
\\
Then condition \eqref{16} may be rewritten as

\begin{equation}
   \label{18} \mathcal{H}_{\mu \nu} := \xi^\rho \partial_\rho \eta_{\mu \nu} + 2 \partial_{(\mu} \xi^\rho \eta_{\nu)\rho} + 2 \varphi \partial_{(\mu} \xi^\rho k_{\nu)}k_\rho \stackrel{!}{=} \zeta(x) k_\mu k_\nu.
\end{equation}
\\
Since $k_\mu$ is a constant vector in spherical Minkowski coordinates, terms containing its derivatives vanish. Thus,

\begin{equation}
   \label{19} 2 \partial_{(\mu} \xi^\rho k_{\nu)} k_\rho = (\partial_\mu \xi^\rho) k_\nu k_\rho + (\partial_\nu \xi^\rho) k_\mu k_\rho.
\end{equation}
\\
Plugging this into \eqref{18} gives a set of ten coupled, nonlinear PDEs for the components of $\xi^\mu$:

\begin{equation}
   \label{20} \mathcal{H}_{\mu \nu} := \xi^\rho \partial_\rho \eta_{\mu \nu} + 2 \partial_{(\mu} \xi^\rho \eta_{\nu)\rho} + \varphi (\partial_\mu \xi^\rho) k_\nu k_\rho + \varphi (\partial_\nu \xi^\rho) k_\mu k_\rho \stackrel{!}{=} \zeta(x) k_\mu k_\nu.
\end{equation}
\\
In spherical coordinates, this system naturally decomposes into angular, radial-temporal, and mixed components, summarized in \textbf{Tables 1-3}.

\begin{table}[h]
    \centering
    \begin{tabular}{|l|l|}
        $\mathcal{H}_{\vartheta \vartheta} $ & $ \xi^r + r \partial_\vartheta \xi^\vartheta \stackrel{!}{=} 0$
        \\
        $\mathcal{H}_{\phi \phi} $ & $ \partial_\vartheta \xi^\vartheta-\xi^\vartheta \cot\vartheta-\partial_\phi \xi^\phi \stackrel{!}{=} 0$
        \\
        $\mathcal{H}_{\vartheta \phi} $ & $ \sin^2\vartheta \partial_\vartheta \xi^\phi + \partial_\phi \xi^\vartheta \stackrel{!}{=} 0$
    \end{tabular}
    \caption{Angular subsystem governing the dependence of $\xi^\mu$ on $(\vartheta,\phi)$ and encoding the conformal Killing structure of the two-sphere.}
    \label{tab:placeholder}
\end{table}

\begin{table}[h]
    \centering
    \begin{tabular}{|l|l|}
        $\mathcal{H}_{tt}$ & $2(1-\varphi) \partial_t \xi^t + 2 \varphi \partial_t \xi^r \stackrel{!}{=} -\zeta(x)$
        \\
        $\mathcal{H}_{rr}$ & $2(1+\varphi)\partial_r \xi^r-2 \varphi \partial_r \xi^t \stackrel{!}{=} \zeta(x)$
        \\
        $\mathcal{H}_{tr} $ & $(1-\varphi) \partial_r \xi^t + \varphi \partial_t \xi^t \stackrel{!}{=} \zeta(x)$
    \end{tabular}
    \caption{Radial-temporal subsystem constraining the $(t,r)$-dependence of $\xi^\mu$.}
    \label{tab:placeholder}
\end{table}

\begin{table}[h]
    \centering
    \begin{tabular}{|l|l|}
        $\mathcal{H}_{t\vartheta}$ & $r^2 \partial_t \xi^\vartheta-(1-\varphi) \partial_\vartheta \xi^t \stackrel{!}{=} 0$
        \\
        $\mathcal{H}_{t\phi}$ & $r^2 \sin^2 \vartheta \partial_t \xi^\phi-(1-\varphi) \partial_\phi \xi^t \stackrel{!}{=} 0$
        \\
        $\mathcal{H}_{r \vartheta}$ & $r^2 \partial_r \xi^\vartheta + \varphi \partial_\vartheta \xi^t \stackrel{!}{=} 0$
        \\
        $\mathcal{H}_{r \phi}$ & $r^2 \sin^2 \vartheta \partial_r \xi^\phi + \varphi \partial_\phi \xi^t \stackrel{!}{=} 0$
    \end{tabular}
    \caption{Mixed subsystem coupling angular and radial-temporal components of $\xi^\mu$ through compatibility conditions.}
    \label{tab:placeholder}
\end{table}
\noindent
We note that $\zeta(x)$ appears only in the radial-temporal subsystem. This reflects the tensor structure of the KS-preserving conditions, since contributions proportional to $k_\mu k_\nu $ arise solely in the $(t,r)$ sector. As a result, the angular equations decouple from $\zeta(x)$ and instead constrain the intrinsic angular dependence of the vector field $\xi^\mu$. With this structure in place, we now proceed to solve the system sector by sector.

\subsection{The Angular Subsystem: Symmetries of the Two-Sphere}

We begin with the angular components of the KS-preserving condition, summarized in \textbf{Table 1}. These PDEs can be written more succinctly as $\mathcal{H}_{AB} = 0$ for $A,B \in \{ \vartheta, \phi \}$. Using the metric decomposition $g_{AB} = r^2 \gamma_{AB}$, the angular equations may be written covariantly as

\begin{equation}
   \label{21} \nabla_A \xi_B + \nabla_B \xi_A =-\frac{2 \xi^r}{r} \gamma_{AB},
\end{equation}
\\
where $\gamma_{AB} := d\vartheta^2 + \sin^2\vartheta d\phi^2$ is the standard metric on the unit two-sphere and $\nabla_A$ is its Levi-Civita connection. Since the right-hand side is proportional to $\gamma_{AB}$, equation \eqref{21} is precisely the conformal Killing equation on $S^2$, with conformal factor $-2\xi^r/r$.
\\ \\
The general solution is well known: conformal Killing vectors on the two-sphere form a six-dimensional space that decomposes uniquely into rotational Killing vectors and proper conformal Killing vectors (CKVs) \cite{r}. Accordingly, the angular components of $\xi^\mu$ take the general form

\begin{equation}
   \label{22} \xi^A(t,r,\vartheta,\phi) = \sum_{i=1}^3 a_i(t,r) \xi_{(i)}^A (\vartheta, \phi) + \sum_{i=1}^3 b_i(t,r) K_{(i)}^A (\vartheta, \phi),
\end{equation}
\\
where $\xi_{(i)}^A$ generate rotations on $S^2$ and $K_{(i)}^A$ denote the proper CKVs. Here, $a_i(t,r)$ and $b_i(t,r)$ are smooth functions. Choosing as our basis the standard generators of rotations about the Cartesian axes, we have,

\begin{equation}
    \begin{split}
    \label{23} \xi_{(1)}^A(\vartheta, \phi) &= (-\sin\phi, -\cot\vartheta \cos\phi), \\ \xi_{(2)}^A(\vartheta, \phi) &= (\cos\phi, -\cot\vartheta \sin\phi), \\ \xi_{(3)}^A(\vartheta, \phi) &= (0, 1).
    \end{split}
\end{equation}
\\
Moreover, the proper CKVs can be written as

\begin{equation}
    \begin{split}
    \label{24}    K_{(1)}^A &= (\cos\vartheta \cos\phi, -\sin\phi/\sin\vartheta), \\
        K_{(2)}^A &= (\cos\vartheta \sin\phi, \cos\phi/\sin\vartheta), \\
        K_{(3)}^A &= (- \sin\vartheta, 0).
    \end{split}
\end{equation}
\\
Since the defining residual symmetry condition is linear in $\xi^\mu$, the Killing and proper conformal Killing sectors decouple, and each may be analyzed independently. To simplify the discussion, we follow this scheme in Sections 3.3 and 3.4.

\subsection{Killing Vectors in the Kerr-Schild Double Copy}

Restricting to the Killing sector of the angular solutions, we set $b_i(t,r) = 0$. In this case, the angular components $\xi^A$ reduce to linear combinations of the rotational Killing vectors on the two-sphere,

\begin{equation}
    \begin{split}
    \label{25}    \xi^\vartheta(t,r,\vartheta,\phi) &=-a_1(t,r) \sin\phi + a_2(t,r) \cos\phi, \\
        \xi^\phi(t,r,\vartheta,\phi) &=-a_1(t,r) \cot\vartheta \cos\phi-a_2(t,r) \cot\vartheta \sin\phi + a_3(t,r),
    \end{split}
\end{equation}
\\
where coefficients $a_i(t,r)$ remain to be determined by the remaining constraints given in \textbf{Tables 2-3}.
\\ \\
From \eqref{25}, $\xi^\vartheta$ is independent of $\vartheta$, so $\mathcal{H}_{\vartheta \vartheta}$ implies

\begin{equation}
   \label{26} \xi^r \stackrel{!}{=}-r \partial_\vartheta \xi^\vartheta = 0.
\end{equation}
\\
The radial-temporal equations, summarized in \textbf{Table 2}, together with condition \eqref{26} constrain the time component $\xi^t$. Solving $\mathcal{H}_{tt}$ for $\partial_t \xi^t$ and $\mathcal{H}_{rr}$ for $\partial_r \xi^t$, respectively, then substituting into $\mathcal{H}_{tr}$ yields the following constraint

\begin{equation}
    \begin{bmatrix}
    \label{27} \frac{(1-\varphi)}{2 \varphi} + \frac{\varphi}{2(1-\varphi)} + 1 
    \end{bmatrix} \zeta(x) = 0,
\end{equation}
\\
which vanishes if and only if $\zeta(x) = 0$. Thus, we find that $\partial_t \xi^t = \partial_r \xi^t = 0$, so $\xi^t$ is independent of $(t,r)$.
\\ \\
The mixed equations, given in\textbf{ Table 3}, place further constraints on $\xi^t$. The reasoning here is subtle, so we explicitly work this out for the reader. First, differentiating $\mathcal{H}_{\vartheta \phi}$ with respect to $t$ and utilizing the fact that partial derivatives commute yields:

\begin{equation}
  \label{28}  \partial_t \partial_\vartheta \xi^\phi \sin^2 \vartheta + \partial_t \partial_\phi \xi^\vartheta = \partial_\vartheta (\partial_t \xi^\phi) \sin^2\vartheta + \partial_\phi (\partial_t \xi^\vartheta) = 0.
\end{equation}
\\
From $\mathcal{H}_{t\vartheta}$ and $\mathcal{H}_{t\phi}$, we know the explicit forms of $\partial_t \xi^\phi$ and $\partial_t \xi^\vartheta$. Substituting into \eqref{28} gives

\begin{equation}
   \label{29} \sin^2\vartheta \partial_\vartheta \begin{pmatrix} \sin^{-2} \vartheta \partial_\phi \xi^t \end{pmatrix} + \partial_\phi \begin{pmatrix} \partial_\vartheta \xi^t \end{pmatrix} = 0.
\end{equation}
\\
By the Leibniz rule, this simplifies to

\begin{equation}
   \label{30} [ \partial_\vartheta-\cot\vartheta ] \partial_\phi \xi^t = 0.
\end{equation}
\\
This can be solved via substitution. Let $g(\vartheta,\phi) := \partial_\phi \xi^t $. Then,

\begin{equation}
   \label{31} g'(\vartheta,\phi)-\cot\vartheta g(\vartheta,\phi) = 0,
\end{equation}
\\
which has general solution

\begin{equation}
   \label{32} g(\vartheta,\phi) = C(\phi) \sin\vartheta
\end{equation}
\\
for smooth function $C(\phi)$. Plugging \eqref{32} into $\mathcal{H}_{t\phi}$, differentiating with respect to $\vartheta$, and noting that $\xi^\vartheta$ is independent of $\vartheta$, we find that $\partial_\vartheta \xi^t$ is also independent of $\vartheta$. Thus, we are free to write

\begin{equation}
   \label{33} \partial_\vartheta \xi^t = A(\phi),
\end{equation}
\\
where $A(\phi)$ is a smooth function. Differentiating $g(\vartheta,\phi)$ with respect to $\vartheta$, we find:

\begin{equation}
   \label{34} \partial_\vartheta g(\vartheta,\phi) = \partial_\vartheta \partial_\phi \xi^t = \partial_\phi \partial_\vartheta \xi^t = \partial_\phi A(\phi).
\end{equation}
\\
This is clearly independent of $\vartheta$. However, $g(\vartheta,\phi)$ depends explicitly on $\vartheta$ unless $C(\phi) = 0$, so it must be the case that $C(\phi) = 0$.
\\ \\
Substituting this result into $\mathcal{H}_{t\vartheta}$ and $\mathcal{H}_{t\phi}$, we find that $\partial_t \xi^\vartheta = \partial_\vartheta \xi^t = 0$ and $\partial_t \xi^\phi = \partial_\phi \xi^t = 0$. Consequently, $\xi^t$ is independent of $(t, r, \vartheta, \phi)$, so $\xi^t$ is constant:

\begin{equation}
  \label{35}  \xi^t = c_1.
\end{equation}
\\
This identifies time translations as residual symmetries in the Killing sector, as expected for Schwarzschild. Furthermore, $\xi^\vartheta$ and $\xi^\phi$ are independent of $(t,r)$. The coefficients $a_i(t,r)$ are, therefore, constants:
\begin{equation}
   \label{36} a_i = \text{constant}.
\end{equation}
\\
Collecting these results, we find that the residual symmetries preserving the KS ansatz in the Killing sector take the form

\begin{equation}
   \label{37} \xi^\mu = \begin{pmatrix} c_1, 0, \xi^A(\vartheta,\phi)\end{pmatrix},
\end{equation}
\\
where $\xi^A(\vartheta,\phi)$ with $A \in \{\vartheta,\phi\}$ correspond precisely to rotations on $S^2$:

\begin{equation}
    \begin{split}
    \label{38}    \xi^\vartheta(\vartheta,\phi) &= -a_1 \sin\phi + a_2 \cos\phi, \\
        \xi^\phi(\vartheta,\phi) &= -a_1 \cot\vartheta \cos\phi-a_2 \cot\vartheta \sin\phi + a_3.
    \end{split}
\end{equation}
\\
These vectors satisfy $\mathcal{H}_{AB} = 0$ and coincide with the Killing vectors of the round two-sphere, and are therefore isometries of $S^2$.
\\ \\
To verify that solutions \eqref{37} and \eqref{38} generate global isometries of the full metric, note that $\zeta(x) = 0$ implies $\alpha(x) = \xi^\rho \partial_\rho \varphi$. Since $\varphi = \varphi(r)$ and $\xi^r = 0$, it follows that $\alpha(x) = 0$, so $(\mathcal{L}_\xi g)_{\mu \nu} = 0$. Hence, \eqref{37} and \eqref{38} are indeed isometries of KS-Schwarzschild spacetime.
\\ \\
We have shown that the Killing sector yields global isometries of the Schwarzschild spacetime: time translations and spatial rotations, which generate the expected, finite-dimensional $\mathfrak{so}(3) \oplus \mathbb{R}$ symmetry algebra under the Lie bracket.
\\ \\
Finally, we remark that because the KS-preserving condition yields a closed and formally integrable system of PDEs, these solutions exhaust all residual symmetries in the Killing class.

\subsection{Proper CKVs in the Kerr-Schild Double Copy}

We now analyze the proper conformal Killing vectors (CKVs). Setting $a_i(t,r) = 0$ and substituting \eqref{24} into \eqref{22} yields

\begin{equation}
    \begin{split}
     \label{AX}  \xi^\vartheta(t,r,\vartheta,\phi) &= b_1(t,r) \cos\vartheta \cos\phi + b_2(t,r) \cos\vartheta \sin\phi-b_3(t,r) \sin\vartheta, \\
        \xi^\phi(t,r,\vartheta,\phi) &=-b_1(t,r) \frac{\sin\phi}{\sin\vartheta} + b_2(t,r) \frac{\cos\phi}{\sin\vartheta},
    \end{split}
\end{equation}
\\
where coefficients $b_i(t,r)$ are determined by the remaining constraints.

\subsubsection{Determining the Angular Dependence of $\xi^t$}

To determine $\xi^t$ and $\xi^r$, note that $\mathcal{H}_{\vartheta \vartheta}$ requires

\begin{equation}
   \label{AY} \xi^r \stackrel{!}{=}-r \partial_\vartheta \xi^\vartheta = -r \textbf{b}(t,r) \cdot \textbf{n}(\vartheta, \phi),
\end{equation}
\\
where $\textbf{b} = (b_1, b_2, b_3)$ and $\textbf{n}(\vartheta,\phi) := (\sin\vartheta \cos\phi, \sin\vartheta \sin\phi, \cos\vartheta)$ are the Cartesian embedding coordinates of the round two-sphere. Thus, the proper CKV sector naturally organizes into the dipole basis $\textbf{n}(\vartheta,\phi)$.
\\ \\
With the angular and radial dependence fixed, the remaining equations arise from the  $tt$, $rr$, $tr$, $tA$, and $rA$ components of $\mathcal{H}_{\mu \nu}$.
\\ \\
In particular, a key simplification follows from the mixed equations $\mathcal{H}_{tA}$ and $\mathcal{H}_{rA}$. Differentiating $\mathcal{H}_{tA}$ with respect to $r$ and $\mathcal{H}_{rA}$ with respect to $t$, taking the difference, and exploiting the commutativity of partial derivatives allows us to determine the angular dependence of $\xi^t$ independently of its remaining coordinates:

\begin{equation}
   \label{AE} \begin{bmatrix} \partial_r \partial_t-\partial_t \partial_r \end{bmatrix} \xi^\vartheta = \begin{bmatrix}-\varphi'(r)-\frac{2(1-\varphi)}{r} + (1-\varphi) \partial_r-\varphi \partial_t \end{bmatrix} \partial_A \xi^t = 0.
\end{equation}
\\
Let $u_A := \partial_A \xi^t$ and

\begin{equation}
    A(r) := -\varphi'(r)-\frac{2(1-\varphi(r))}{r} ~~~~~,~~~~~B(r) := 1-\varphi(r) ~~~~~,~~~~~C(r) := -\varphi(r).
\end{equation}
\\
Then, \eqref{AE} can be written schematically as

\begin{equation}
    \begin{bmatrix}
        B(r) \partial_r + C(r) \partial_t 
    \end{bmatrix} u_A =-A(r) u_A.
\end{equation}
\\
This equation is solved via the method of characteristics. Introducing characteristic curves labeled by

\begin{equation}
    \chi := t + \int^r \frac{\varphi(\rho) }{1-\varphi(\rho)} d\rho,
\end{equation}
\\
we find that the general solution takes the form

\begin{equation}
  \label{AF} u_A := \partial_A \xi^t = H_A(\chi, \vartheta, \phi) \exp \begin{bmatrix}
       -\int^r \frac{A(\rho)}{B(\rho)} d\rho
    \end{bmatrix},
\end{equation}
\\
where $H_A(\vartheta,\phi)$ encodes all the residual freedom in $\partial_A \xi^t$ for $A \in \{\vartheta, \phi\}$ while remaining constant along the characteristic.
\\ \\
The integrability condition $\partial_\phi u_\vartheta = \partial_\vartheta u_\phi$ ensures that both components derive from a single scalar function $H(\chi, \vartheta,\phi)$ so that $H_A(\chi, \vartheta,\phi) = \partial_A H(\chi, \vartheta,\phi)$. Reconstructing $\xi^t$ gives

\begin{equation}
    \xi^t(t,r,\vartheta,\phi) = \exp \begin{bmatrix}-\int^r \frac{A(\rho)}{B(\rho)} d\rho \end{bmatrix} H(\chi, \vartheta,\phi) + S(t,r),
\end{equation}
\\
where $S(t,r)$ is an integration function determined by the remaining equations.
\\ \\
For Schwarzschild, the radial integral evaluates to

\begin{equation}
    \exp \begin{bmatrix}-\int^r \frac{A(\rho)}{B(\rho)} d\rho \end{bmatrix} \propto \frac{r^3}{r-2GM},
\end{equation}
\\
with additional scale factors absorbed into $H(\chi,\vartheta,\phi)$. The general temporal component becomes

\begin{equation}
   \label{AZ} \xi^t(t,r,\vartheta,\phi) = \frac{r^3}{r-2GM} H(\chi,\vartheta,\phi) + S(t,r).
\end{equation}
\\
At this stage, $H(\chi,\vartheta,\phi)$ is completely arbitrary. However, the structure of the remaining PDEs restricts its angular content. Because all angular dependence entering the remaining PDEs appears only through the constant mode and the dipole basis $\textbf{n}(\vartheta,\phi)$, higher spherical harmonics decouple from the system. Accordingly, we expand

\begin{equation}
    H(\chi,\vartheta,\phi) = Q(\chi) + \textbf{P}(\chi) \cdot \textbf{n}(\vartheta,\phi),
\end{equation}
\\
where $Q(\chi)$ and $\textbf{P}(\chi)$, which has components $P_i(\chi)$, represent the monopole and dipole modes, respectively. Substituting into \eqref{AZ} gives

\begin{equation}
  \label{AXZ} \xi^t(t,r,\vartheta,\phi) = \frac{r^3}{r-2GM} \begin{bmatrix} Q(\chi) + \textbf{P}(\chi) \cdot \textbf{n}(\vartheta,\phi) \end{bmatrix} + S(t,r).
\end{equation}

\subsubsection{Integrability and the Radial Coefficients $b_i(t,r)$}

The remaining constraints arise from the $\mathcal{H}_{tt}$, $\mathcal{H}_{rr}$, and $\mathcal{H}_{tr}$ equations. Combining them to eliminate $\zeta(x)$ yields the remarkably simple integrability condition

\begin{equation}
   (\partial_t + \partial_r)(\xi^t-\xi^r) = 0,
\end{equation}
\\
which follows directly from the structure of the radial-temporal subsystem. Let $D:=\partial_t + \partial_r$ so that the integrability condition reads $D \xi^t = D \xi^r$, giving,

\begin{equation}
    \begin{split}
      \label{ABC}  -D [r\textbf{b}(t,r)] \cdot \textbf{n}(\vartheta,\phi) &= \begin{bmatrix} F'(r) \textbf{P}(\chi) + \frac{F(r)}{1-\varphi(r)} \textbf{P}'(\chi) \end{bmatrix} \cdot \textbf{n}(\vartheta,\phi) \\ &+ F'(r) Q(\chi) + \frac{F(r)}{1-\varphi(r)} Q'(\chi) + DS(t,r),
    \end{split}
\end{equation}
\\
where $F(r) := \frac{r^3}{r-2GM}$. Here, $F'(r) = dF/dr$ while $Q'(\chi) = dQ/d\chi$ and $\textbf{P}'(\chi) = d\textbf{P}/d\chi$. For any $\chi$-dependent scalar $G(\chi)$, we remark that 

\begin{equation}
   \label{40} DG(\chi) = \frac{G'(\chi)}{1-\varphi(r)}.
\end{equation}
\\
\textbf{Note:} the left-hand side is strictly proportional to $\textbf{n}(\vartheta,\phi)$, and is a dipole term. Conversely, the right-hand side contains both monopole and dipole terms. Consistency requires cancellation of the monopole terms. This fixes $DS(t,r)$ uniquely as

\begin{equation}
  \label{41}  DS(t,r) :=-\begin{bmatrix}
        F'(r) Q(\chi) + \frac{F(r)}{1-\varphi(r)} Q'(\chi)
    \end{bmatrix}.
\end{equation}
\\
Substituting into \eqref{ABC}, we obtain a PDE for $r\textbf{b}(t,r)$ that is completely independent of the angular coordinates. Namely,

\begin{equation}
  \label{42} D[r\textbf{b}(t,r)] =-\begin{bmatrix} F'(r) \textbf{P}(\chi) + \frac{F(r)}{1-\varphi(r)} \textbf{P}'(\chi) \end{bmatrix}.
\end{equation}
\\
Let $\textbf{B}(t,r) := r \textbf{b}(t,r)$, so that we have

\begin{equation}
  \label{43} D\textbf{B} =-\begin{bmatrix} F'(r) \textbf{P}(\chi) + \frac{F(r)}{1-\varphi(r)} \textbf{P}'(\chi) \end{bmatrix}.
\end{equation}
\\
The transport equation is first-order along the vector field $D$. Its characteristics satisfy $dt/dr = 1$, so that $u=t-r$ is constant along each characteristic. Along these curves, we parametrize the solution by $r$, so that $D = d/dr$ and the equation reduces to

\begin{equation}
   \label{44} \frac{d\textbf{B}}{dr} =-\begin{bmatrix} F'(\rho) \textbf{P}(\chi) + \frac{F(\rho)}{1-\varphi(\rho)} \textbf{P}'(\chi)\end{bmatrix}.
\end{equation}
\\
Integrating from an arbitrary reference radius, $r_0$, then restoring the original coordinates gives

\begin{equation}
   \label{45} \textbf{B}(t,r) = \textbf{B}_0(u)-\int_{r_0}^r \begin{pmatrix} F'(\rho) \textbf{P}(\chi) + \frac{F(\rho)}{1-\varphi(\rho)} \textbf{P}'(\chi)\end{pmatrix} d\rho.
\end{equation}
\\
Replacing $\textbf{B}(t,r) = r \textbf{b}(t,r)$ yields the following integral equation for $\textbf{b}(t,r)$:

\begin{equation}
   \label{AZZ} \textbf{b}(t,r) = \frac{1}{r} \begin{bmatrix} \textbf{B}_0(u)-\int_{r_0}^r \begin{pmatrix} F'(\rho) \textbf{P}(\chi) + \frac{F(\rho)}{1-\varphi(\rho)} \textbf{P}'(\chi)\end{pmatrix} d\rho \end{bmatrix},
\end{equation}
\\
Together, \eqref{AX}, \eqref{AY}, \eqref{AXZ}, and \eqref{AZZ} give the most general formal solution for the proper CKV sector preserving the KS form. After enforcing consistency of the radial-temporal equations, the solution is parameterized by three independent functions along outgoing null directions: the monopole mode $Q(\chi)$, the dipole terms $\textbf{P}(\chi)$, and the integration functions $\textbf{B}_0(u)$.

\subsection{Asymptotic Flatness}

We now impose asymptotic flatness by examining the proper CKV sector in the limit $r \rightarrow \infty$. Using the general solution derived in Section 3.4, we analyze the leading radial behavior of the coefficients. In the asymptotic region (at fixed $t$), one has

\begin{equation}
    F(r) \sim r^{2} ~~~~~,~~~~~\chi \sim \ln(r-2GM) ~~~~~,~~~~~\varphi(r) \sim r^{-1}.
\end{equation}
\\
See \eqref{QAO} below for the full expression of $\chi$. The integral term appearing in the solution for $\textbf{b}(t,r)$ then behaves as

\begin{equation}
    \int_{r_0}^r \begin{bmatrix}
        F'(\rho) \textbf{P}(\chi) + \frac{F(\rho)}{1-\varphi(\rho)} \textbf{P}'(\chi)
    \end{bmatrix} d\rho \sim \int_{r_0}^r \begin{bmatrix}
        2 \rho \textbf{P}(\chi) + \rho^2 \textbf{P}'(\chi)
    \end{bmatrix} d\rho.
\end{equation}
\\
Unless $\textbf{P}(\chi) = 0$, this grows at least quadratically in $r$, so $\textbf{b}(t,r) \sim r$, and the corresponding vector field diverges at spatial infinity. Asymptotic flatness requires residual diffeomorphisms to remain bounded, which is satisfied only if

\begin{equation}
    \textbf{P}(\chi) = 0.
\end{equation}
\\
Hence, the radial coefficients reduce to

\begin{equation}
    \textbf{b}(t,r) = \frac{1}{r} \textbf{B}_0(u).
\end{equation}
\\
Thus, asymptotic flatness completely removes the dipole sector while leaving the monopole mode unconstrained.

\subsection{Horizon Regularity}

Having imposed asymptotic flatness, we now examine the behavior of the proper CKV sector near the Schwarzschild horizon $r = 2GM$ to enforce regularity. After asymptotic reduction, the remaining solution is characterized by the monopole modes $Q(\chi)$, $S(t,r)$, and $\textbf{B}_0(u)$
\\ \\
Near the horizon, the characteristic variable behaves as

\begin{equation}
   \label{QAO} \chi = t + \int^t \frac{2GM}{\rho-2GM} d\rho \sim t + 2GM \ln(r-2GM) + \text{finite}.
\end{equation}
\\
Hence, $\chi$ diverges logarithmically as $r \rightarrow 2GM$. From the general solution,

\begin{equation}
    \xi^t = \frac{r^3}{r-2GM} Q(\chi) + S(t,r).
\end{equation}
\\
Unless $Q(\chi) = 0$, the first term diverges at the horizon. Regularity therefore requires

\begin{equation}
    Q(\chi) = 0,
\end{equation}
\\
and the temporal component reduces to $\xi^t(t,r) = S(t,r)$. The remaining KS-preserving conditions then impose the transport constraint

\begin{equation}
    DS(t,r) = (\partial_t + \partial_r)S(t,r) = 0,
\end{equation}
\\
so that

\begin{equation}
    S(t,r) = S(t-r) = S(u),
\end{equation}
\\
a smooth function of the outgoing null coordinate $u$.
\\ \\
The radial component retains its monopole form along null characteristics,

\begin{equation}
    \xi^r(t,r,\vartheta,\phi) = -\textbf{B}_0(u) \cdot \textbf{n}(\vartheta,\phi),
\end{equation}
\\
while the angular components follow from the CKV structure on the round two-sphere:

\begin{equation}
    \begin{split}
        \xi^\vartheta(t,r,\vartheta,\phi) &= \frac{B_{0,1}(u)}{r} \cos\vartheta \cos\phi + \frac{B_{0,2}(u)}{r} \cos\vartheta \sin\phi-\frac{B_{0,3}(u)}{r} \sin\vartheta, \\
        \xi^\phi(t,r,\vartheta,\phi) &=-\frac{B_{0,1}(u)}{r} \frac{\sin\phi}{\sin\vartheta} + \frac{B_{0,2}(u)}{r} \frac{\cos\phi}{\sin\vartheta},
    \end{split}
\end{equation}
\\
where $B_{0,i}$ are the components of $\textbf{B}_0(u)$. Thus, horizon regularity removes the remaining monopole $Q(\chi)$ while restricting the surviving functions to null dependence.
\\ \\
The emergence of null dependence is natural in this setting. Both asymptotic flatness and horizon regularity constrain the residual diffeomorphisms to propagate along the outgoing null direction generated by $k_\mu$, so that the surviving functions depend only on $u = t-r$. This mirrors the gauge theory case, where KS-preserving residual gauge transformations are likewise constant along null characteristics.

\subsection{Algebras Generated by the Residual Diffeomorphisms}

To determine the algebra of residual CKV symmetries, we compute the Lie bracket of two CKV generators $\eta_1^\mu$ and $\eta_2^\mu$. We use this notation for the remainder of Section 3 to distinguish Killing vectors, which we maintain as $\xi$, from the proper CKVs. After imposing asymptotic flatness and horizon regularity, the residual vector fields take the form

\begin{equation}
   \label{AHAHA} \eta^\mu = \begin{pmatrix}
        S(u), -\textbf{B}(u) \cdot \textbf{n}(\vartheta,\phi), \eta^A(u,\vartheta,\phi)
    \end{pmatrix},
\end{equation}
\\
where $u = t-r$, $\textbf{n}(\vartheta,\phi)$ is the dipole basis on the unit two-sphere, and the angular components $\eta^A$ are fixed by the CKV structure. Since all coefficient functions depend only on $u$, derivatives reduce to $\partial_t = \partial_u$ and $\partial_r =-\partial_u$.
\\ \\
A direct computation shows that the commutator again takes the form of a residual vector field. For the temporal component, we find

\begin{equation}
   \label{AHA} [\eta_1,\eta_2]^t = \eta_1^\mu \partial_\mu S_2-\eta_2^\mu \partial_\mu S_1 = (S_1 + \textbf{B}_1 \cdot \textbf{n}) S_2'-(S_2 + \textbf{B}_2 \cdot \textbf{n}) S_1',
\end{equation}
\\
where primes denote derivatives with respect to $u$. Here, $S_i(u)$ are scalar functions, $\textbf{B}_i(u)$ are three-component functions valued in the dipole basis on $S^2$, and $\textbf{n}(\vartheta,\phi)$ denotes the unit vector on $S^2$.
\\ \\
Similarly, the radial component yields

\begin{equation}
   \label{AHAH} [\eta_1,\eta_2]^r =-\begin{bmatrix}
        (S_1 + \textbf{B}_1 \cdot \textbf{n}) \textbf{B}_2'
    -(S_2 + \textbf{B}_2 \cdot \textbf{n}) \textbf{B}_1' \end{bmatrix} \cdot \textbf{n},
\end{equation}
\\
where the result is understood as a projection onto the radial direction defined by $\textbf{n}(\vartheta,\phi)$.
\\ \\
The angular components close automatically, as they are determined by the conformal Killing structure on $S^2$ with $u$-dependent coefficients. Introducing the shorthand

\begin{equation}
    \Sigma_i(u,\vartheta,\phi) := S_i(u) + \textbf{B}_i(u) \cdot \textbf{n}(\vartheta,\phi),
\end{equation}
\\
commutators \eqref{AHA} and \eqref{AHAH} can be written compactly as

\begin{equation}
        [\eta_1,\eta_2]^t = S_{12}(u,\vartheta,\phi) ~~~~~,~~~~~
        [\eta_1,\eta_2]^r = -B_{12}(u)\cdot \textbf{n}(\vartheta,\phi),
\end{equation}
\\
where,

\begin{equation}
    S_{12} = \Sigma_1 S_2'-\Sigma_2 S_1' ~~~~~,~~~~~ \textbf{B}_{12} = \Sigma_1 \textbf{B}_2'-\Sigma_2 \textbf{B}_1'.
\end{equation}
\\
Note: primes denote derivatives with respect to $u$. Thus, the commutator takes the form

\begin{equation}
    [\eta_1,\eta_2]^\mu = \begin{pmatrix} S_{12}(u), -\textbf{B}_{12}(u) \cdot \textbf{n}, \eta^A_{12} \end{pmatrix},
\end{equation}
\\
where $\eta_{12}^A$ is the angular component determined by $\textbf{B}_{12}(u)$ through the CKV structure on $S^2$. This result has the same functional structure as the residual vector field \eqref{AHAHA}. Hence, the residual transformations are closed under the Lie bracket.

\subsection{Closure of the Algebra}

Before we move forward, it is instructive to verify explicitly that the algebra closes under mixed commutators between the Killing and proper CKV sectors. In this way, all potential \enquote{hidden} symmetries of the Kerr-Schild ansatz are excluded.
\\ \\
Let $\xi$ denote a Killing vector and $\eta$ denote a proper CKV. Using that all coefficient functions depend only on $u$, we find:

\begin{equation}
    \xi^\mu \partial_\mu = c_1 \partial_u + \xi^A \partial_A,
\end{equation}
\\
again for $A \in \{\vartheta, \phi\}$. It is trivial to see that

\begin{equation}
    [\xi, \eta]^t = c_1 S'(u),
\end{equation}
\\
which preserves the scalar structure of the CKV sector. For the radial component, we obtain

\begin{equation}
    [\xi, \eta]^r = -c_1 \textbf{B}'(u) \cdot \textbf{n} - \textbf{B}(u) \cdot (\xi^A \partial_A \textbf{n}).
\end{equation}
\\
The second term arises from angular derivatives acting on the dipole basis. Using that Killing vectors generate rotations on $S^2$, we obtain

\begin{equation}
    \xi^A \partial_A n_i = \Omega_{ij} n_j,
\end{equation}
\\
where $\Omega_{ij} = -\Omega_{ji}$ is an antisymmetric matrix generating an element of $\mathfrak{so}(3)$, determined by the rotational Killing vector $\xi^A$. This contribution, therefore, remains of dipole form:

\begin{equation}
    [\xi, \eta]^r = - \begin{bmatrix} c_1 \textbf{B}'(u) + \Omega \textbf{B}(u)\end{bmatrix} \cdot \textbf{n}.
\end{equation}
\\
The angular components close by the conformal Killing algebra on $S^2$, which is preserved pointwise in $u$. Therefore, the mixed Lie bracket preserves the structure of the residual vector fields, and we have shown that

\begin{equation}
   [\eta_1, \eta_2 ] \in \mathfrak{g}_{\text{CKV}} ~~~~~,~~~~~ [\xi, \eta] \in \mathfrak{g}_{\text{CKV}}
\end{equation}
\\
for $\xi \in \mathfrak{g}_{\text{iso}} $ and $\eta_1, \eta_2, \eta \in \mathfrak{g}_{\text{CKV}}$.
\\ \\
Additionally, the Killing sector closes separately to the finite-dimensional algebra $\mathfrak{so}(3) \oplus \mathbb{R}$ under the Lie bracket,

\begin{equation}
   [\xi_1, \xi_2 ] \in \mathfrak{g}_{\text{iso}},
\end{equation}
\\
for $\xi_1,\xi_2 \in \mathfrak{g}_{\text{iso}}$.
\\ \\
Together with the closure of the proper CKV and mixed sectors, this shows that the gravitational residual symmetries form a closed infinite-dimensional algebra of null-dependent vector fields.
\\ \\
Unlike the Yang-Mills case, where residual transformations form the current algebra $\mathfrak{g} \otimes C^\infty(\mathbb{R})$, the gravitational bracket involves derivatives with respect to $u$ and mixes the null-dependent coefficients nonlinearly. Moreover, since no additional structures are generated under commutation, the algebra does not enlarge beyond the class of vector fields already identified. We therefore conclude that, subject to the Kerr-Schild preservation condition together with asymptotic flatness and horizon regularity, the residual symmetries derived above exhaust the allowed transformations. In this sense, there are no hidden residual symmetries within this ansatz, and the full residual symmetry algebra takes the form of a semidirect product

\begin{equation}
    \mathfrak{g}_{\text{res}} \cong \mathfrak{g}_{\text{iso}} \ltimes \mathfrak{g}_{\text{CKV}},
\end{equation}
\\
with the isometry algebra acting on the null-dependent CKV sector by derivations along $u$ and rotations of the dipole coefficients.

\section{BRST Cohomology and Physical Symmetry Reduction}

In the previous sections, we derived the full set of residual diffeomorphisms preserving the Kerr-Schild structure of the Schwarzschild solution. These decompose into two distinct sectors:

\begin{itemize}
    \item the Killing sector, generating the finite-dimensional isometry algebra

    \begin{equation}
        \mathfrak{g}_{\text{iso}} \cong \mathfrak{so}(3) \oplus \mathbb{R}.
    \end{equation}

    \item the proper conformal Killing vector (CKV) sector, forming an infinite-dimensional algebra $\mathfrak{g}_{\text{CKV}}$ parametrized by arbitrary functions $h(u) =\{ S(u), \textbf{B}(u)\}$.
\end{itemize}
\noindent
The full residual symmetry algebra is therefore $\mathfrak{g}_{\text{res}} \cong \mathfrak{g}_{\text{iso}} \ltimes \mathfrak{g}_{\text{CKV}}.$
\\ \\
While this enlarged algebra is a genuine feature of the Kerr-Schild representation, it is not isomorphic to the symmetry algebra of canonical Schwarzschild spacetime. In this section, we show that this apparent mismatch is resolved cohomologically: the infinite-dimensional CKV sector is entirely BRST-exact, and the physical symmetry algebra reduces to the finite-dimensional isometries.

\subsection{BRST in the Killing Sector}

The residual symmetries derived above act on fields via Lie derivatives. However, they do not in general define a representation of the full residual symmetry algebra $\mathfrak{g}_{\text{res}}$ on the field space: in particular, the proper conformal Killing vectors close only up to Weyl rescalings, as we show in the next section. We therefore begin by constructing the BRST complex for the Killing sector, where the algebra acts faithfully, and subsequently extend the construction to the full residual symmetry algebra.
\\ \\
Restricting to the Killing sector $\mathfrak{g}_{\text{iso}} \in \mathfrak{g}_{\text{res}}$, the action on any field $\Psi$ is given by

\begin{equation}
    \delta_\epsilon \Psi = \epsilon^a \mathcal{L}_{K_a} \Psi ~~~~~,~~~~~ K_a \in \mathfrak{g}_{\text{iso}}.
\end{equation}
\\
Promoting the infinitesimal parameters $\epsilon^a$ to Grassmann-odd ghosts $c^a$, we define the BRST operator:

\begin{equation}
    \mathcal{Q}_K \Psi = c^a \mathcal{L}_{K_a} \Psi ~~~~~,~~~~~ \mathcal{Q}_K c^a = -\frac{1}{2} {f_{bc}}^a c^b c^c,
\end{equation}
\\
where ${f_{bc}}^a$ are the structure constants of $\mathfrak{g}_{\text{iso}}$. 
\\ \\
Nilpotency, $\mathcal{Q}_K^2 = 0$, follows directly from the Jacobi identity.
\\ \\
For the Schwarzschild background in Kerr-Schild form, the Killing vectors satisfy

\begin{equation}
    \mathcal{L}_{K_a} g_{\mu \nu} = 0,
\end{equation}
\\
and similarly leave all Kerr-Schild fields invariant. It follows that the representation of $\mathfrak{g}_{\text{iso}}$ on the field space is trivial,

\begin{equation}
    \mathcal{Q}_{K} \Psi = 0.
\end{equation}
\\
The BRST differential therefore reduces to its purely algebraic part,

\begin{equation}
    \mathcal{Q}_{K} c^a = -\frac{1}{2} {f_{bc}}^a c^b c^c,
\end{equation}
\\
so that the complex computes the Chevalley-Eilenberg (CE) cohomology of $\mathfrak{g}_{\text{iso}}$ with trivial coefficients. No nontrivial BRST variations of the physical fields arise in this sector.
\\ \\
Thus, the BRST complex takes the form

\begin{equation}
    \mathcal{C}^{\bullet} = \mathcal{F} \otimes \Lambda^{\bullet} \mathfrak{g}_{\text{iso}}^*.
\end{equation}
\\
Here, $\mathcal{F}$ denotes the space of fields, while $\Lambda^{\bullet} \mathfrak{g}_{\text{iso}}^*$ is the exterior algebra generated by the ghost variables $c^a$, so that the BRST complex consists of fields valued in polynomials of ghosts, graded by ghost number.
\\ \\
Hence, $\mathcal{Q}_K$ computes the Lie algebra cohomology of the isomorphism algebra $\mathfrak{g}_{\text{iso}}$.
\\ \\
The form of $\mathcal{Q}_K$ is not an additional assumption, but follows uniquely from the Lie algebra structure of $\mathfrak{g}_{\text{iso}}$. In particular, the BRST differential coincides with the Chevalley-Eilenberg differential computing Lie algebra cohomology with trivial coefficients. Equivalently, the Killing sector reproduces the standard BRST complex for global symmetries.

\subsection{BRST in the Proper CKV Sector}

We now turn to the proper conformal Killing vector (CKV) sector. By definition, a conformal Killing vector $\Xi^\mu$ satisfies

\begin{equation}
    \mathcal{L}_{\Xi} g_{\mu \nu} = \nabla_\mu \Xi_\nu + \nabla_\nu \Xi_\mu = \Omega(x;h) g_{\mu \nu},
\end{equation}
\\
where the conformal factor $\Omega(x;h)$ is not an independent function, but is determined by the divergence of the vector field,

\begin{equation}
    \Omega(x;h) = \frac{2}{d} \nabla_\mu \Xi^\mu.
\end{equation}
\\
Thus, the action of a CKV on the metric is fixed entirely by the generator $\Xi_h$, and does not preserve the metric, but instead produces a local Weyl rescaling. As a result,

\begin{equation}
    \mathcal{L}_{\Xi_h} g_{\mu\nu} = \Omega(x;h) g_{\mu \nu}.
\end{equation}
\\
Thus, the CKV transformations do not preserve the metric, but instead generate local Weyl rescalings. As a result, they do not define a representation of the symmetry algebra on the space of fields $\mathcal{F}$, and the standard BRST construction does not directly apply.
\\ \\
To obtain a well-defined BRST complex, it is therefore necessary to enlarge the field space so that the symmetry acts faithfully. This is achieved by introducing a scalar compensator field $\Phi$, which transforms under Weyl rescalings. The CKV transformations can then be interpreted as combined diffeomorphisms and Weyl transformations acting on the extended field space.
\\ \\
Promoting the CKV parameters $h(u)$ to Grassmann-odd ghost fields $c(u)$, we define

\begin{equation}
    \mathcal{Q}_W \Phi = \Omega(x;c),
\end{equation}
\\
so that the compensator absorbs the conformal variation. The BRST transformation of the metric is then defined by combining the diffeomorphism and compensating Weyl transformation,

\begin{equation}
    \mathcal{Q}_W g_{\mu \nu} = \mathcal{L}_{\Xi_c} g_{\mu \nu}-(\mathcal{Q}_W \Phi) g_{\mu \nu}.
\end{equation}
\\
By construction, the conformal variation cancels:

\begin{equation}
    \mathcal{Q}_W g_{\mu \nu} = 0.
\end{equation}
\\
After imposing the Kerr-Schild constraints and regularity conditions, the CKV algebra closes trivially on the extended field space, so that it is effectively Abelian in this sector. Accordingly, the BRST variation of the ghosts reduces to

\begin{equation}
    \mathcal{Q}_W c(u) = 0,
\end{equation}
\\
and nilpotency, $\mathcal{Q}_W^2 = 0$, follows immediately.
\\ \\
Thus, in contrast to the Killing sector, the CKV transformations become trivial in the extended field space. The introduction of the compensator ensures that the CKV sector defines a consistent BRST complex, but one in which all transformations act trivially on physical fields.
\\ \\
In this sense, the proper CKV sector defines a BRST-contractible subcomplex, and therefore does not contribute to the cohomology.

\subsection{Unified BRST Complex and Cohomological Reduction}

We now combine both sectors into a single BRST operator

\begin{equation}
    \mathcal{Q}_{\text{tot}} = \mathcal{Q}_K + \mathcal{Q}_W.
\end{equation}
\\
The extended field space is

\begin{equation}
    \Psi_{\text{tot}} = \{ g_{\mu \nu}, \Phi, c^a, c(u) \}.
\end{equation}
\\
By construction,

\begin{equation}
    \mathcal{Q}_{\text{tot}} g_{\mu \nu} = 0 ~~~~~,~~~~~ \mathcal{Q}_{\text{tot}}^2 = 0,
\end{equation}
\\
so the Kerr-Schild metric is BRST-closed.
\\ \\
The physical content of the theory is encoded in the BRST cohomology

\begin{equation}
    H(\mathcal{Q}_{\text{tot}}) = \frac{\ker(\mathcal{Q}_{\text{tot}})}{\text{im}(\mathcal{Q}_{\text{tot}})}.
\end{equation}
\\
The structure of the cohomology can be determined without an explicit computation of $\ker(\mathcal{Q}_{\text{tot}})$ and $\text{im}(\mathcal{Q}_{\text{tot}})$ by exploiting the decomposition of the BRST complex into independent sectors.
\\ \\
From Section 4.1, the Killing sector defines a standard Chevalley-Eilenberg complex with trivial action on the fields. The corresponding generators are BRST-closed and not exact, and therefore contribute nontrivial cohomology classes.
\\ \\
From Section 4.2, the proper CKV sector becomes trivial in the extended field space: the metric and all physical fields are invariant under $\mathcal{Q}_W$, while the ghost sector satisfies $\mathcal{Q}_W c(u) = 0$.
\\ \\
Moreover, the compensator field $\Phi$ and the conformal factor $\Omega(x;c)$ form a BRST doublet, whose BRST variation reproduces the conformal factor,

\begin{equation}
    \mathcal{Q}_W \Phi = \Omega(x;c) ~~~~~,~~~~~\mathcal{Q}_W \Omega(x;c) = 0.
\end{equation}
\\
Such pairs are cohomologically trivial: any functional depending on them is either BRST-exact or equivalent to one that does not depend on them. This implies that the CKV sector defines a BRST-contractible subcomplex, and hence, does not contribute to the cohomology.
\\ \\
It follows that the full BRST cohomology reduces to that of the Killing sector,

\begin{equation}
    H(\mathcal{Q}_{\text{tot}}) \cong H(\mathcal{Q}_K) \cong \mathfrak{so}(3) \oplus \mathbb{R}.
\end{equation}
\\
Equivalently, the unified BRST complex implements a cohomological projection

\begin{equation}
    \mathfrak{g}_{\text{res}} \rightarrow H(\mathcal{Q}_{\text{tot}}) \cong \mathfrak{g}_{\text{iso}},
\end{equation}
\\
recovering precisely the finite-dimensional isometry algebra of Schwarzschild spacetime.

\subsection{Physical Interpretation}

The Kerr-Schild ansatz introduces an enlarged residual symmetry algebra

\begin{equation}
     \mathfrak{g}_{\text{res}} =  \mathfrak{g}_{\text{iso}} \ltimes  \mathfrak{g}_{\text{CKV}},
\end{equation}
\\
arising from the null structure of the decomposition. The infinite-dimensional CKV sector reflects this additional geometric structure.
\\ \\
However, BRST cohomology identifies these modes as pure gauge. The Weyl compensator absorbs the conformal variation, rendering the CKV sector entirely BRST-exact. These transformations do not correspond to physical degrees of freedom, but instead represent redundancies of the Kerr-Schild representation.

\subsection{Summary}

We have shown that:

\begin{itemize}
    \item The Kerr-Schild residual symmetry algebra is enlarged by an infinite-dimensional CKV sector.
    \item The associated BRST complex is the Chevalley-Eilenberg complex of the full symmetry algebra.
    \item The CKV sector becomes BRST-exact upon introducing a minimal Weyl compensator.
    \item The BRST cohomology reduces the symmetry algebra to the physical isometries.
\end{itemize}
\noindent
This establishes that, although the Kerr-Schild double copy does not preserve residual symmetry algebras at the geometric level, it remains fully consistent at the level of physical observables.

\section{Discussion}

In this work, we have investigated the structure of residual symmetries in the Kerr-Schild (KS) double copy for the Schwarzschild solution, with particular emphasis on their algebraic and cohomological properties. Our analysis reveals a clear and instructive separation between the symmetry structure at the level of the KS ansatz and the physical symmetry algebra of the underlying spacetime.
\\ \\
A central result of this paper is that the residual symmetry algebra preserving the KS form of the Schwarzschild metric is significantly enlarged relative to the isometry algebra of the spacetime. In addition to the expected finite-dimensional Killing sector $\mathfrak{g}_{\text{iso}} \cong \mathfrak{so}(3) \oplus \mathbb{R}$, we find an infinite-dimensional sector generated by proper conformal Killing vectors (CKVs), parametrized by arbitrary functions along null directions. This enhancement originates from the intrinsic null structure of the Kerr-Schild decomposition and has no direct analogue in the canonical formulation of the Schwarzschild solution.
\\ \\
At first sight, this result appears to be in tension with the standard understanding of Schwarzschild spacetime, which admits no proper conformal symmetries. The resolution of this apparent discrepancy is provided by the BRST framework developed in Section 4. By constructing a unified BRST complex adapted to the residual symmetry algebra, we have shown that the infinite-dimensional CKV sector is entirely BRST-exact. The introduction of a minimal Weyl compensator allows the CKV transformations to be realized consistently on an extended field space, but simultaneously renders them cohomologically trivial. As a result, the physical symmetry algebra is recovered as the BRST cohomology,

\begin{equation}
    H(\mathcal{Q}_{\text{tot}}) \cong \mathfrak{g}_{\text{iso}},
\end{equation}
\\
in agreement with the expected isometries of Schwarzschild spacetime.
\\ \\
Conceptually, this provides a natural interpretation of the enlarged symmetry algebra: the additional CKV modes do not correspond to genuine physical symmetries, but instead reflect redundancies inherent to the Kerr-Schild representation. In this sense, the KS ansatz introduces a form of “parametrization gauge freedom,” whose associated transformations are removed by BRST cohomology. This perspective clarifies how an infinite-dimensional symmetry structure can arise at the level of the ansatz without modifying the physical content of the theory.
\\ \\
These results have important implications for the Kerr-Schild double copy program. While the double copy is often viewed as relating solutions of Yang-Mills theory and gravity, our analysis shows that this correspondence does not extend straightforwardly to the level of residual symmetry algebras. In particular, the Yang-Mills residual symmetries form a current algebra $\mathfrak{g} \otimes C^\infty(\mathbb{R})$, whereas the gravitational residual symmetries exhibit a more intricate structure involving nonlinear mixing of null-dependent functions. The fact that the gravitational CKV sector is ultimately BRST-trivial suggests that any putative symmetry-level double copy must be formulated at the level of cohomology, rather than at the level of raw symmetry algebras.
\\ \\
From this perspective, our results may help clarify the role of geometric symmetry structures in the classical double copy more broadly. Recent work has shown that certain gauge-theoretic sectors admit diffeomorphism and kinematic algebra structures associated with symplectic or area-preserving transformations \cite{fstar}. The residual symmetry algebras studied in the present work arise in a similar fashion, but from geometric structures tied directly to the Kerr-Schild decomposition, suggesting that diffeomorphism-type algebras may constitute a recurring feature of classical double copy constructions. At the same time, our analysis demonstrates that enlarged symmetry structures appearing at the level of a particular representation need not survive as physical symmetries after cohomological reduction. In this sense, BRST cohomology provides a natural framework for distinguishing physically meaningful symmetry data from representation-dependent redundancies within broader symmetry-based approaches to the double copy.
\\ \\
Additionally, the appearance of contractible sectors and cohomological reductions is familiar in gauge theory, but its role in the context of Kerr-Schild geometries and the double copy has not been widely explored. It would be interesting to investigate whether similar structures arise in more general spacetimes admitting Kerr-Schild representations, such as Kerr or AdS backgrounds, and whether the interplay between conformal symmetries and BRST triviality persists in those cases.
\\ \\
Several further directions suggest themselves. First, one may ask whether the cohomological reduction observed here admits a direct interpretation on the gauge theory side of the double copy. Second, it would be natural to explore whether the Weyl-compensated BRST construction can be embedded into a more general BV framework \cite{z, kk, xa}, where the role of auxiliary fields and contractible pairs can be treated systematically. Finally, understanding how these structures interact with asymptotic symmetries \cite{o, oa, bab, hh}, hidden symmetries \cite{ha}, and soft theorems \cite{oa, baa} may shed further light on the symmetry foundations of the double copy.
\\ \\
In summary, we have shown that the apparent enhancement of residual symmetries in the Kerr-Schild formulation of Schwarzschild spacetime is a representation-level effect, which is removed upon passing to BRST cohomology. This provides a concrete example in which the double copy preserves physical symmetries only after an appropriate cohomological reduction, and underscores the importance of distinguishing between geometric and physical symmetry structures in the study of classical solutions.




\end{document}